\documentclass[prd,reprint,showpacs,showkeys,twocolumn]{revtex4}
\usepackage{textcomp}
\usepackage{eurosym}
\usepackage{amsfonts}
\usepackage{array}
\usepackage{amsthm}
\usepackage{bm}
\usepackage[colorinlistoftodos]{todonotes}
\usepackage{mathpazo}
\usepackage{supertabular}
\usepackage{subfig}
\usepackage{amssymb}
\usepackage{eurosym}
\usepackage{amsmath}
\usepackage{epsfig}
\usepackage{graphics}
\usepackage{color}
\usepackage{graphicx}
\usepackage[font={footnotesize,it}]{caption}
\usepackage[colorlinks=true,
            linkcolor=blue,
            urlcolor=blue,
            citecolor=green]{hyperref}
\usepackage{braket}
\usepackage{amsmath}
\makeatother

\begin{document}

\title{Hawking Radiation in the Spacetime of White Holes}

\author{Kimet Jusufi}
\email{kimet.jusufi@unite.edu.mk}
\affiliation{Physics Department, State University of Tetovo, Ilinden Street nn, 1200,
Tetovo, Macedonia} 
\affiliation{Institute of Physics, Faculty of Natural Sciences and Mathematics, Ss. Cyril
and Methodius University, Arhimedova 3, 1000 Skopje, Macedonia}

\affiliation{Physics Department, State University of Tetovo, Ilinden Street nn,
1200, Macedonia}

\date{\today }
\begin{abstract}
A white hole (WH) is a time-reversed black hole (BH) solution in General relativity with a spacetime region to which cannot be entered from the outside. Recently they have been proposed as a possible solution to the information loss problem [Haggard and Rovelli, 2015].  In particular it has been argued that the quantization of the gravitational field may prevent a BH from collapsing entirely with an exponential decay law associated to the black-hole-to-white-hole (BHWH) tunneling scenario [Barcelo, Carballo, and Garay, 2017]. During this period of BHWH transition the Hawking radiation should take place. Taking this possibility into account, we utilize the Hamilton-Jacobi and Parkih-Wilczek  methods to study the Hawking radiation viewed as a quantum tunneling effect to calculate the tunneling rate of vector particles tunneling inside (outside) the horizon of a WH (BH), respectively.  We show that there is a Hawking radiation associated to a WH spacetime equal to the BH Hawking temperature when viewed from the outside region of the WH geometry.  In the framework of Parkih-Wilczek method, surprisingly, we show that Hawking temperature is affected by the initial radial distance at which the gravitational collapse starts.
\end{abstract}


\maketitle
%\tableofcontents
\section{Introduction}

Black holes  undoubtedly have attracted and continue to attracted great interest among the physicst. Even thought they were predicted 100 years ago by Einstein's theory of relativity, historically, black holes were met with wide-spread scepticism. Today, however, the situation has changed substantially due to the indirect experimental data supporting their existence.  In the context of classical general relativity, black holes are thought as a region of spacetime that nothing can escape from inside it-not even light. Forty years ago, Stephen Hawking changed this view. He showed that black holes should radiate particles due to the quantum effects near the event horizon \cite{hawking1,hawking2,hawking3}. Hawking radiation is a well understood phenomenon which has been extensively studied in the past. One such an interesting method is the quantum tunneling method where Hawking temperature can be found by simply estimating  the tunneling rate of particles tunneling outside the black hole \cite{kraus1,kraus2,perkih1,perkih2,perkih3,t3,t4,t5,t6,t7,majhi,t8,t9,t10,
t11,t12,t13,t14,t15,t16,t17,t18,t19,t20,t21,t22,t23,t24,t25,t26,t27,t28,t29,t30,
t31,t32}. 

Althoughthe the mathematics of Hawking radiation is shown to be correct by several methods, it was realized by Hawking himself that a serious conceptual problem arises due to the black hole evaporation known as the information loss problem. To solve this problem, several solutions have been proposed over the years. In a recent work by Haggard and Rovelli \cite{rovelli} a new possibility of releasing the information form the black hole has been proposed. This idea involves a BHWH tunneling scenario and certainly remains as an open possibility which has yet to be achieved by some quantum gravity theory.  It is speculated that when gravitational collapse reaches the Planckian scale, a quantum bounce may take place leading to the BHWH tunneling \cite{kiefer}. It is speculated that a WH acts as a long-lived remnant, in this way WHs can be shed lighton the black-hole information paradox \cite{r3,r4,r5,r6}. White holes are of primary phenomenological interest in the analogue gravity, in this context Hawking radiation from dispersive theories in acoustic white holes is studied in Ref. \cite{r1}, while acoustic white holes in flowing atomic Bose-Einstein condensates has been studied in Ref. \cite{r2}. From the astrophysical point of view, high energy radiation from white holes was studied in Ref. \cite{r22}.

In a very interesting work \cite{barcelo1}, authors have put forward this idea by considering a path integral approach they found an exponential law associated to the BHWH tunneling probability. However, the framework used by the authors is highly idealized and not free from ambiguity, for example, they have assumed that the spacetime is static before and after the bounce. Yet there is a major problem from the physical point of view concerning the BHWH transition due to the second law of thermodynamics. Namely, as we know, the gravitational collapse can lead to a black hole with an enormous amount of entropy. Contrary to this, a possible WH formation should be followed by a decrease of entropy which makes their existence very unlikely in nature. During the evolution of BHWT quantum transition the Hawking effect has not been taken into account. Although this effect is negligible, nevertheless, it will be interesting to study this effect in the context of WH spacetime which is the main purpose of this paper.

This paper is  organised as follows. In Sec. II, we review the generalized Painlev\'{e} coordinates \cite{kanai} and the BHWH tunneling considered in Ref. \cite{barcelo1}.  In Sec. III, we shall solve the Proca equation to find the radial motion of massive vector particles using Hamilton-Jacobi (HJ) method.  In Sec. IV, we study Hawking radiation in the spacetime of a BH. In Sec. V, we shall focus on the Hawking radiation in the spacetime of a WH. In Sect. VI, we study the tunneling of scalar particles. In Sect. VII we solve this problem in the framework of Parkih-Wilczek (PW) method. Finally, in Sect. VIII we present our conclusions.

\section{Black-hole-to-white-hole}
Let us start by briefly reviewing a bouncing spacetime geometry outside a collapsing gravitational process which can be given by a time-symmetric bounce in terms of the following metric 
\cite{barcelo1,barcelo2,barcelo3,kanai} 
\begin{equation}\label{1}
\mathrm{d}s^{2}=-\mathrm{d}t^{2}+\frac{\left(\mathrm{d}r-f(u)v(r)\mathrm{d}t\right)^2}{1-2M/r_i}+r^2\left(\mathrm{d}\theta^2+\sin^2\theta \, \mathrm{d}\varphi ^2\right),
\end{equation}
provided $r_i>2M$. It is worth noting that the metric \eqref{1} represents a generalized Painlev\'{e} type metric which is different from the standard Painlev\'{e} metric. There are two particular cases, $f(u)=\pm 1 $, corresponding to the WH (BH) spacetimes, respectively. The function $v(r)$ is given by 
\begin{equation}\label{2}
v(r)=\left(\frac{2M}{r}-\frac{2M}{r_i}\right)^{1/2},\,\,\,\,\, r_{s}(t)<r<r_i.
\end{equation}

In the above equation, $r_i$ gives the initial radial distance at which the gravitational collapse starts. One should keep in mind that the value of $r_i$ is far above the Schwarzschild radius. Furthermore $r_s(t)$ represents the trajectory of the surface of the star $r_s(t)$ at a given time $t$. In the BH case, from a classical point of view (singularity theorems) it is well known that such a collapse ends with a singularity. 
Here, in contrast, the singularity  $r_s(t) = 0$ is not achieved. One way to achieve this is to simply continue  $r_s(t)$ with it's time-reversal solution with the spacetime geometry to which cannot be entered from the outside i.e., a WH geometry. 

From a quantum mechanical point of view, the amplitude between two configurations, say $h_{-}$ and $h_{+}$, with the corresponding hypersurfaces $\Sigma_{-}$ and $\Sigma_{+}$, related to the BHWH tunneling is given by  \cite{barcelo1}
\begin{equation}
\braket{h_{+},\Sigma_{+}|h_{-},\Sigma_{-}}=\frac{1}{\mathcal{N}}\int_{{g(\Sigma_{-})=h_{-}}}^{{g(\Sigma_{+})=h_{+}}} \mathcal{D}g\,\exp\left(-\mathcal{L}_{EH}[g]\right),
\end{equation}
where $\mathcal{L}_{EH}[g]$ is the Einstein-Hilbert action of a Euclidean geometry $g$, while $\mathcal{N}$ being a normalization constant. Then the probability amplitude for the BHWH transition can be given as \cite{barcelo1}
\begin{equation}
\mathcal{P}_{BH \to WH}(M, \Delta_0)=\int_0^{\Delta_0}|\braket{WH|BH}_{M,\Delta^{\prime}_0}|^2 \,\mathrm{d}\Delta^{\prime}_0,
\end{equation}
where $k_{I}\in [1,3]$ and $\Delta_0 \in [0,\infty)$. In Ref. \cite{barcelo1}, the authors were able to estimate the BHWH transition probability given by an exponential decay law 
\begin{equation}
\mathcal{P}_{BH \to WH}(M, \Delta_0)=1-\exp\left(- \frac{2(k_{I}+1/3)M \Delta_0}{\sqrt{1-2M/r_i}}   \right).
\end{equation}

Then taking into account that $k_{I} + 1/3$ is of the order of unity, the mean lifetime was found to be $\tau \leq 1/2M$.  \\

\section{Tunneling of Vector particles with HJ method}
In order to proceed to study the quantum tunneling let us first introduce a new radial function  given by
\begin{equation}
\mathcal{F}(r)=1-\frac{2M}{r},
\end{equation}
which after we substitute into equation \eqref{2} gives 
\begin{equation}
v(r)=\left(1-\mathcal{F}(r)-\frac{2M}{r_i}\right)^{1/2}.
\end{equation}

More specifically metric \eqref{1} has the following metric tensor components 
\begin{equation}
g_{\mu\nu}=\begin{bmatrix}
\vspace{0.3cm}
\frac{r_i-2M-f^2(u)v^2(r)r_i}{2M-r_i} & \frac{f(u) v(r) r_i}{2M-r_i} & 0 & 0 \\
\vspace{0.3cm}
\frac{f(u) v(r) r_i}{2M-r_i} & \frac{r_i}{r_i-2M} & 0 & 0\\
\vspace{0.3cm}
0 & 0 & r^2 & 0 \\
\vspace{0.3cm}
0 & 0 & 0 & r^2 \sin^2\theta
\end{bmatrix},
\end{equation}
with the determinant given by
\begin{equation}
g=\det \left( g_{\mu\nu}\right)=- \frac{r_i r^4 \sin^2\theta}{r_i-2M}.
\end{equation}

We shall study the tunneling of massive vector particles described by the vector field $\Psi^{\mu}$, in a curved spacetime metric given by the Proca equation (PE) as follows \cite{t11}
\begin{eqnarray}
\frac{1}{\sqrt{-g}}\partial_{\mu}\left(\sqrt{-g}\,\Psi^{\mu\nu}\right)-\frac{m^{2}}{\hbar^{2}}\Psi^{\nu}=0,
\end{eqnarray}
 with the following relation
\begin{equation}
\Psi_{\mu\nu}=\partial_{\mu}\Psi_{\nu}-\partial_{\nu}\Psi_{\mu}.
\end{equation}

One way to solve PE  is to apply the WKB approximation method with the solution proposed as
\begin{equation}\label{12}
\Psi_{\nu}(x^{\mu})=C_{\nu}(x^{\mu})\exp\left(\frac{i}{\hbar}\left(S_{0}(x^{\mu})+\hbar\,S_{1}(x^{\mu})+\hdots.\right)\right).
\end{equation}

We now take take into consideration the spacetime symmetries which lead to the following ansatz for the action of the particle
\begin{equation}
S_{0}(t,r,\theta,\varphi)=-E\,t+R(r,\theta)+j\varphi,
\end{equation}
with $E$ being the energy, and $j$  being the angular momentum of the particle. Inserting Eq. \eqref{12} into the Proca equation and keeping only the leading order of $\hbar$ results with the following four differential equations:

\begin{eqnarray}\notag
0&=&\left[\frac{Er^2\sin^2\theta\left(2M-r_i  \right)R^{\prime}(r)-r_i\mathcal{K}_1 }{r_i r^2 \sin^2 \theta}\right]C_1\\\notag
&+&\left[ \frac{f v (\partial_\theta R) R^{\prime} -E (\partial_\theta R) }{r^2}\right]C_2+\left[\frac{f v R^{\prime} j -E j}{r^2 \sin^2\theta}  \right]C_3\\
&+& \left[\frac{\sin^2\theta (2M-r_i)r^2 R^{\prime 2}-r_i \mathcal{K}_2}{r_i r^2 \sin^2\theta}   \right]C_4,
\end{eqnarray}
\begin{eqnarray}\notag
0&=&\left[\frac{-(f^2v^2r_i+2M-r_i)(\partial_\theta R)^2 \sin^2\theta-r^2 \mathcal{K}_3+\mathcal{K}_4}{r_i r^2 \sin^2\theta}\right]C_1 \\\notag
&+& \left[  \frac{(f^2 v^2 r_i+2M-r_i)(\partial_\theta R) R^{\prime}-r_i f v E (\partial_\theta R)}{r_i r^2}\right]C_2\\
&+&\left[\frac{(f^2 v^2 j r_1+(2M-r_i)j) R^{\prime}-Ef v j r_i}{r^2 r_i \sin^2 \theta}  \right]C_3 \\\notag
&+& \left[ \frac{-r_i \sin^2\theta f v (\partial_\theta R)^2+(2M-r_i) r^2 E \sin^2\theta +\mathcal{K}_5}{r_i^2 r^2 \sin^2\theta} \right]C_4,
\end{eqnarray}
\begin{eqnarray}\notag
0&=&\left[\frac{(f^2 v^2 r_i+2M-r_i)(\partial_\theta R) R^{\prime}-r_i f v E (\partial_\theta R)}{r_i r^2}\right]C_1 \\\notag
&+& \left[ \frac{-r^2 \sin^2\theta (f^2 v^2 r_i^2+2M-r_i)R^{\prime 2}+r_i\mathcal{K}_6}{r_i r^4 \sin^2\theta}  \right]C_2\\\notag
&+&\left[ - \frac{(\partial_\theta R) j}{r^4 \sin^2\theta}\right]C_3 \\
&+& \left[ \frac{f v (\partial_\theta R) R^{\prime} -E (\partial_\theta R) }{r^2}\right]C_4,
\end{eqnarray}
\begin{eqnarray}\notag
0&=&\left[ \frac{(f^2 v^2 j r_1+(2M-r_i)j) R^{\prime}-Ef v j r_i}{r^2 r_i \sin^2 \theta} \right]C_1\\\notag
&+& \left[ - \frac{(\partial_\theta R) j}{r^4 \sin^2\theta} \right]C_2 \\\notag
&+& \left[\frac{-r^2  (f^2 v^2 r_i^2+2M-r_i)R^{\prime 2}+2 Ef v r_i r^2 R^{\prime}-\mathcal{K}_7}{r_i r^4 \sin^2\theta}\right]C_3 \\
&+& \left[ \frac{f v R^{\prime} j -E j}{r^2 \sin^2\theta}\right]C_4.
\end{eqnarray}
where we have used the following relations
\begin{eqnarray}\notag
\mathcal{K}_1 &=&f v \sin^2\theta  (\partial_\theta R)^2+m^2r^2f v \sin^2\theta+j^2 vf ,\\\notag
\mathcal{K}_2&=&\sin^2\theta (\partial_\theta R)^2+m^2 r^2 \sin^2\theta +j^2,\\\notag
\mathcal{K}_3&=&\cos^2\theta (f^2 v^2 m^2 r_i+(2M-r_i)(E^2-m^2)),\\\notag
\mathcal{K}_4&=&(2M-r_i)((E^2-m^2)r^2-j^2)-v^2f^2r_i (m^2r^2+j^2),\\\notag
\mathcal{K}_5&=& fv m^2 r^2 r_i \cos^2\theta - fvr_i (m^2r^2+j^2), \\\notag
\mathcal{K}_6&=& 2 Ef v  r^2 \sin^2\theta R^{\prime}-[(E^2-m^2)r^2 \sin^2\theta -j^2],\\\notag
\mathcal{K}_7&=&r_i[(E^2-m^2)r^2 -(\partial_\theta R)^2].
\end{eqnarray}

Resulting with the following non-zero matrix elements
\begin{eqnarray}\notag
\mathbb{M}_{11} & = & \frac{Er^2\sin^2\theta\left(2M-r_i  \right)R^{\prime}(r)-r_i\mathcal{K}_1 }{r_i r^2 \sin^2 \theta}, \\\notag
\mathbb{M}_{21} & = & \frac{-(f^2v^2r_i+2M-r_i)(\partial_\theta R)^2 \sin^2\theta-r^2 \mathcal{K}_3+\mathcal{K}_4}{r_i r^2 \sin^2\theta} ,\\\notag
\mathbb{M}_{31} & = &\mathbb{M}_{22}=  \frac{(f^2 v^2 r_i+2M-r_i)(\partial_\theta R) R^{\prime}-r_i f v E (\partial_\theta R)}{r_i r^2}, \\\notag
\mathbb{M}_{41} & = & \mathbb{M}_{23}  =  \frac{(f^2 v^2 j r_1+(2M-r_i)j) R^{\prime}-Ef v j r_i}{r^2 r_i \sin^2 \theta},
\end{eqnarray}
\begin{eqnarray}\notag
\mathbb{M}_{12} & = & \mathbb{M}_{34}= \frac{f v (\partial_\theta R) R^{\prime} -E (\partial_\theta R) }{r^2},\\\notag
\mathbb{M}_{32} & = &  \frac{-r^2 \sin^2\theta (f^2 v^2 r_i^2+2M-r_i)R^{\prime 2}+r_i\mathcal{K}_6}{r_i r^4 \sin^2\theta}, \\\notag
\mathbb{M}_{42} & = & \mathbb{M}_{33}= - \frac{(\partial_\theta R) j}{r^4 \sin^2\theta}, \\\notag
\mathbb{M}_{13}  & = & \mathbb{M}_{44}=\frac{f v R^{\prime} j -E j}{r^2 \sin^2\theta}, \\\notag
\mathbb{M}_{14}  & = & \frac{\sin^2\theta (2M-r_i)r^2 R^{\prime 2}-r_i \mathcal{K}_2}{r_i r^2 \sin^2\theta} ,\\\notag
\mathbb{M}_{24}  & = & \frac{-r_i \sin^2\theta f v (\partial_\theta R)^2+(2M-r_i) r^2 E \sin^2\theta +\mathcal{K}_5}{r_i^2 r^2 \sin^2\theta},\\\notag
\mathbb{M}_{43}  & = & \frac{-r^2  (f^2 v^2 r_i^2+2M-r_i)R^{\prime 2}+2 Ef v r_i r^2 R^{\prime}-\mathcal{K}_7}{r_i r^4 \sin^2\theta}.
\end{eqnarray}

The radial motion of the particles is easily found if we solve the determinant 
\begin{equation}
\det \mathbb{M}(C_{1},C_{2},C_3,C_4)^{T}=0,
\end{equation}
which yields 
\begin{equation}
\frac{m^2 (2M-r_i)\left[\mathcal{H}-2 E R^{\prime} v r^2 r_i f \sin^2\theta - r_i \mathcal{G} \right]^3}{r^{10} r_i^4 \sin^8\theta}=0,
\end{equation}
with
\begin{eqnarray}\notag
\mathcal{H} &=& r^2 \sin^2\theta (f^2v^2 r_i+2M-r_i)R^{\prime 2},\\\notag
\mathcal{G} &=&\sin^2\theta (\partial_{\theta}R)^2-(E^2-m^2)r^2 \sin^2\theta +j^2.
\end{eqnarray}

Finally, after solving for the radial trajectories results with
\begin{equation}\label{20}
R_{\pm}(r)=\int\frac{E f(u)v(r) r_i \pm \sqrt{E^2r_i^2 \left(1-\frac{2M}{r_i}\right)-\Delta \,\mathcal{N} }}{r_i \Delta(r) } \mathrm{d}r,
\end{equation}
with
\begin{eqnarray}
\Delta &=& v^2(r)f^2(u)-\left(1-\frac{2M}{r_i}\right),\\
\mathcal{N}&=&\left[m^2+\frac{(\partial_{\theta}R)^2 r_i^2 }{r^2}+\frac{j^2 r_i^2 }{r^2 \sin^2\theta}\right].
\end{eqnarray}
\bigskip

\section{Quantum tunneling in a BH geometry}
Let us now consider the quantum tunneling in the spacetime of a BH [$f(u)=-1$]. The function $\Delta$ can be written as 
\begin{equation}
\Delta=\left(-v(r)-\sqrt{1-\frac{2M}{r_i}}\right)\left(-v(r)+\sqrt{1-\frac{2M}{r_i}}\right). 
\end{equation}

Near the horizon we may expand this function which yields
\begin{eqnarray}
\Delta(r_h) &\simeq &-2\,\kappa_{BH}(r-r_h)+..., 
\end{eqnarray}
in which we have used the following identification for the surface gravity of the black hole
\begin{equation}
\kappa_{BH}=\frac{\mathcal{F}^{\prime}(r_h)}{2}>0.
\end{equation}

The radial solution \eqref{20} near the horizon reads
\begin{equation}
R_{\pm}(r_h)=\int \frac{-Ev(r_h) r_i \pm \sqrt{E^2r_i^2 \left(1-\frac{2M}{r_i}\right)-\Delta(r_h)\mathcal{N}}}{-2\, r_i \,\kappa_{BH}(r-r_h) }\mathrm{d}r.
\end{equation}

At this point, we can use the following identity
\begin{equation}
\lim_{\epsilon \to 0}\text{Im} \frac{1}{r-r_{h}+ i\epsilon}=\pi \delta (r-r_{h}),
\end{equation}
which lead to a non-zero contribution only for the outgoing solution $R_{-}(r)$ 
\begin{equation}
\text{Im}R_{-}(r)=\frac{ \pi \mathcal{E} }{\kappa_{BH}},\,\,\,\,\text{Im}R_{+}(r)=0.
\end{equation}
where $\mathcal{E}=E\,v(r_h)$, is the net energy of the particle with $v(r_h)=\sqrt{1-2M/r_i}$. Of course, this result is to be expected as a consequence of the coordinate system used in our setup. There is no  barrier experienced  by the ingoing particle across the event horizon. But, clearly, this is not the case for the outgoing particle. In this way if we define the tunneling rate from the inside to the outside 
\begin{equation}
\Gamma_{BH}=\frac{\Gamma_{out}}{\Gamma_{in}}=\frac{\exp\left(-2\text{Im} R_{-}\right)}{\exp\left(-2 \text{Im}
R_{+}\right)}=\exp\left(-\frac{2 \pi \mathcal{E}}{\kappa_{BH}}\right),
\end{equation}
after we compare the tunneling rate with the Boltzmann equation $%
[\Gamma_{B}=\exp(-\mathcal{E}/T_{H})]$, we easily find the Hawking temperature  
\begin{equation}
T_{H}=\frac{\kappa_{BH}}{2 \pi}=\frac{\mathcal{F}^{\prime}(r_h)}{4 \pi}.
\end{equation}

This result was to be expected, except that this procedure is not free from ambiguity. It has been shown that the above method can lead to the factor--two problem \cite{Akhmedova1,borun}. As was pointed out in Refs. \cite{Akhmedova1} the factor--two problem can be solved when we consider the invariance under canonical transformations by considering a closed path we can write
\begin{equation}
\oint p_{r}\mathrm{d}r=\int_{r_i}^{r_f} p_{r}^{in}\mathrm{d}r+\int_{r_f}^{r_i} p_{r}^{out}\mathrm{d}r. 
\end{equation}

Note that the path goes from just outside the horizon, say $r = r_i$, to $r = r_f$ which is located just inside of the horizon. Next, we shall first work out the spatial contribution to the tunneling rate (we temporarily introduce the Planck constant) \cite{Akhmedova1}
\begin{eqnarray}\notag
\Gamma&=&\exp\left(-\frac{1}{\hbar}\text{Im} \oint p_{r}\mathrm{d}r\right)\\
&=&\exp\left[-\frac{1}{\hbar} \text{Im} \left(\int p_{r}^{in}\mathrm{d}r+\int p_{r}^{out}\mathrm{d}r\right)\right],
\end{eqnarray}
with $p_{r}=\partial_{r}R$.  Next one can shift the pole into the upper half plane $r_{h}\to r_{h}+i\epsilon$ and  rewrite the last equation as follows
\begin{widetext}
\begin{eqnarray}\notag
\text{Im} \oint p_{r} \mathrm{d}r&=&\lim_{\epsilon \to 0}\left\lbrace\text{Im}\left[\int_{r_{i}}^{r_{f}} \frac{-Ev(r) r_i + \sqrt{E^2r_i^2 \left(1-\frac{2M}{r_i}\right)-\Delta(r)\left[m^2+\frac{(\partial_{\theta}R)^2 r_i^2 }{r^2}+\frac{j^2 r_i^2 }{r^2 \sin^2\theta}\right]}}{ -2\, r_i \,\kappa_{BH}(r-r_h+i\epsilon) } \mathrm{d}r\right] \right\rbrace\\
&+&\lim_{\epsilon \to 0}\left\lbrace\text{Im}\left[\int_{r_{f}}^{r_{i}} \frac{-Ev(r) r_i - \sqrt{E^2r_i^2 \left(1-\frac{2M}{r_i}\right)-\Delta(r)\left[m^2+\frac{(\partial_{\theta}R)^2 r_i^2 }{r^2}+\frac{j^2 r_i^2 }{r^2 \sin^2\theta}\right]}}{-2\, r_i \,\kappa_{BH}(r-r_h+i\epsilon) } \mathrm{d}r \right]\right\rbrace.
\end{eqnarray}
\end{widetext}

A non-zero contribution gives only the second term leading to
\begin{equation}
\text{Im} \oint p_{r} \mathrm{d}r=\frac{ \pi \mathcal{E}}{\kappa_{BH}}.
\end{equation}

It remains now to evaluate the temporal contribution to the tunneling rate. To do so, we introduce the following coordinate transformation
\begin{equation}
\mathrm{d}t\to \mathrm{d}\bar{t}-\frac{v(r)f(u) \mathrm{d}r}{v^2(r)f^2(u)-\left(1-\frac{2M}{r_i}\right)}.
\end{equation}

In the last equation $t$ corresponds to the Painlev\'{e} time, while $\bar{t}$ corresponds to the time measured by a far-away observer. Combining this relation with the action gives
\begin{equation}
S_0^{BH}=- E\,\bar{t}+\int \frac{v(r_h)E }{2 \,\kappa_{BH}(r-r_h)} \mathrm{d}r+R(r,\theta)+j\varphi,
\end{equation}
resulting with the following contribution
\begin{equation}
\text{Im}(E\Delta t^{out})=\text{Im}(E\Delta t^{in})=\frac{ \pi \mathcal{E}}{2\,\kappa_{BH}}.
\end{equation}

Thus, by putting these contributions into the total tunneling rate we are left with 
\begin{eqnarray}\notag
\Gamma_{BH} &=&\exp \Big[-\frac{1}{\hbar}\big(\text{Im} (E\Delta t^{out})+\text{Im}(E\Delta t^{in})\\
&+& \text{Im }\oint p_{r} \mathrm{d}r\big)\Big]= \exp \left(-\frac{2\pi \mathcal{E} }{\kappa_{BH}}\right).
\end{eqnarray}

This yields the Hawking temperature 
\begin{equation}
T_{H}=\frac{\kappa_{BH}}{2 \pi}=\frac{\mathcal{F}^{\prime}(r_h)}{4\pi}.
\end{equation}

\section{Quantum tunneling in a WH geometry  }

We tern our attention to the quantum tunneling in the spacetime of a WH geometry [$f(u)=1$]. In that case, the function $\Delta$ is written as
\begin{equation}
\Delta(r) =\left(v(r)-\sqrt{1-\frac{2M}{r_i}}\right)\left(v(r)+\sqrt{1-\frac{2M}{r_i}}\right).
\end{equation}

Near the horizon we find
\begin{eqnarray}
\Delta(r_h) &\simeq &2\,\kappa_{WH}(r-r_h)+..., 
\end{eqnarray}
where we have used the following identification for the surface gravity  of the white hole
\begin{equation}
\kappa_{WH}=-\frac{\mathcal{F}^{\prime}(r_h)}{2}<0.
\end{equation}

With this result in mind, the radial part gives
\begin{equation}
R_{\pm}(r_h)=\int \frac{Ev(r_h) r_i \pm \sqrt{E^2r_i^2 \left(1-\frac{2M}{r_i}\right)-\Delta(r_h)\mathcal{N}}}{2\, r_i \,\kappa_{WH}(r-r_h) }\mathrm{d}r,
\end{equation}
resulting with a non-zero contribution only for the ingoing  solution $R_{+}(r)$ 
\begin{equation}
\text{Im}R_{-}(r)=0,\,\,\,\,\text{Im}R_{+}(r)=\frac{ \pi \mathcal{E} }{\kappa_{WH}}.
\end{equation}
where $\mathcal{E}=E\,v(r_h)$ is a net energy of the particle, in which $v(r_h)=\sqrt{1-2M/r_i}$.  

Let us note that there is a crucial difference in contrast with the BH case, namely there is a problem of the definition of asymptotic ingoing flux inside the WH. In particular by considering the analogue WH model, at a classical level, one can argue that any mode transiting from the outside to the inside will be spitted out again because of this infinite valued flow, hence one expects the transmission coefficient to be zero. In other words, only the outgoing mode is of primary interest, thus we can define the tunneling rate from inside to outside the WH as follows 
\begin{equation}
\Gamma_{WH}=\frac{\Gamma_{out}}{\Gamma_{in}}=\frac{\exp\left(-2\text{Im} R_{-}\right)}{\exp\left(-2 \text{Im}
R_{+}\right)}=\exp\left(+\frac{2\pi \mathcal{E}}{\kappa_{WH}}\right).
\end{equation}
After we compare with the Boltzmann distribution law  $%
\Gamma_{B}=\exp(-\mathcal{E}/T_{H})$, we end up with an interesting result
\begin{equation}
T_{H}=-\frac{\kappa_{WH}}{2 \pi}=\frac{\mathcal{F}^{\prime}(r_h)}{4 \pi}.
\end{equation}

Now we shall derive this result in terms of the invariance under canonical transformations. In this case the path goes from just inside the WH horizon, say $r = r_i$, to $r = r_f$, just outside the WH horizon
\begin{equation}
\oint p_{r}\mathrm{d}r=\int_{r_i}^{r_f} p_{r}^{out}\mathrm{d}r+\int_{r_f}^{r_i} p_{r}^{in}\mathrm{d}r. 
\end{equation}

In particular after we shift the pole $r_{h}\to r_{h}-i\epsilon$, we can write
\begin{widetext}
\begin{eqnarray}\notag
\text{Im} \oint p_{r} \mathrm{d}r&=&\lim_{\epsilon \to 0}\left\lbrace\text{Im}\left[\int_{r_{i}}^{r_{f}} \frac{Ev(r_h) r_i + \sqrt{E^2r_i^2 \left(1-\frac{2M}{r_i}\right)-\Delta(r_h)\left[m^2+\frac{(\partial_{\theta}R)^2 r_i^2 }{r^2}+\frac{j^2 r_i^2 }{r^2 \sin^2\theta}\right]}}{2\, r_i \,\kappa_{WH}(r-r_h-i\epsilon) } \mathrm{d}r\right] \right\rbrace\\
&+&\lim_{\epsilon \to 0}\left\lbrace\text{Im}\left[\int_{r_{f}}^{r_{i}} \frac{Ev(r_h) r_i -\sqrt{E^2r_i^2 \left(1-\frac{2M}{r_i}\right)-\Delta(r_h)\left[m^2+\frac{(\partial_{\theta}R)^2 r_i^2 }{r^2}+\frac{j^2 r_i^2 }{r^2 \sin^2\theta}\right]}}{2\, r_i \,\kappa_{WH}(r-r_h-i\epsilon) } \mathrm{d}r \right]\right\rbrace.
\end{eqnarray}
\end{widetext}

To ensure that the positive result for the Hawking temperature we change the direction in which we shift the pole, that is equivalent to change the direction in which we deform the contour. More precisely we shall use the following equation
\begin{equation}
\lim_{\epsilon \to 0}\text{Im} \frac{1}{r-r_{h}- i\epsilon}=-\pi \delta (r-r_{h}),
\end{equation} 

Again, there is a non-zero contribution for the ingoing particle given by the first term. Put it differently, in the WH spacetime, only the ingoing particle experiences barrier across the horizon. The spatial contribution to the tunneling rate gives
\begin{equation}
\text{Im} \oint p_{r} \mathrm{d}r=-\frac{\pi \mathcal{E}}{\kappa_{WH}}.
\end{equation}

It remains to be seen the temporal contribution. Let us introduce the following coordinates
\begin{equation}
\mathrm{d}t\to \mathrm{d}\bar{t}+\frac{v(r) \mathrm{d}r}{v^2(r)f^2(u)-\left(1-\frac{2M}{r_i}\right)}.
\end{equation}

Again, $t$ corresponds to the Painlev\'{e} time, while $\bar{t}$ corresponds to the time measured by a far-away observer outside the WH geometry. The action of the particle gives
\begin{equation}
S_0^{WH}=- E\,\bar{t}-\int \frac{v(r_h)E }{2 \,\kappa_{WH}(r-r_h)} \mathrm{d}r+R(r,\theta)+j\varphi.
\end{equation}

We find
\begin{equation}
\text{Im}(E\Delta t^{out})=\text{Im}(E\Delta t^{in})=-\frac{\pi \mathcal{E}}{\kappa_{WH}}.
\end{equation}

Hence the total tunneling rate is found to be
\begin{eqnarray}\notag
\Gamma_{WH} &=&\exp \Big[-\frac{1}{\hbar}\big(\text{Im} (E\Delta t^{out})+\text{Im}(E\Delta t^{in})\\
&+& \text{Im }\oint p_{r} \mathrm{d}r\big)\Big]= \exp \left(\frac{2\pi \mathcal{E} }{\kappa_{WH}}\right).
\end{eqnarray}

And the expected result reads
\begin{equation}\label{52}
T_{H}=-\frac{\kappa_{WH}}{2\pi}=\frac{\mathcal{F}^{\prime}(r_h)}{4 \pi}.
\end{equation}

This result suggest that an observer  outside the spacetime of a WH should detected Hawking quanta, in other words a flux of radiation from inside to outside the WH. In addition, the equation of Hawking temperature is form invariant to the BH temperature. 

\bigskip

\section{Tunneling of scalar particles}
In this section we shall explore in details the tunneling of spineless particles, namely massive scalar particles. The relativistic scalar field equation can be written as follows
\begin{equation}
\frac{1}{\sqrt{-g}}\partial_{\mu} \left(g^{\mu \nu} \partial_\nu \Phi   \right)-\frac{m^2}{\hbar^2}\Phi=0.
\end{equation}

The scalar wave solution can be chosen as follows 
\begin{equation}
\Phi(t,r,\theta,\varphi)=C(t,r,\theta,\varphi) \exp \left(\frac{i}{\hbar}(S_0(t,r,\theta,\varphi)+...)  \right),
\end{equation}

Solving the KG equation in leading order terms we find the differential equation
\begin{widetext}
\begin{eqnarray}\notag
\frac{1}{r^2 r_i \sin^2\theta}&&\Big[\sin^2\theta (f^2(u)v^2(r) r_i+2M-r_i)r^2 (\partial_r S_0)^2+2r^2f(u)v(r)r_i (\partial_t S_0)(\partial_r S_0)\\
&+&r_i\left(\sin^2\theta r^2 (\partial_t S_0)^2-\sin^2\theta (\partial_{\theta} S_0)^2-(\partial_\varphi S_0)^2-r^2m^2 \sin^2\theta)\right)\Big]=0
\end{eqnarray}
\end{widetext}
with the same action 
\begin{equation}
S_0(t,r,\theta,\varphi)=-Et+R(r,\theta)+j\varphi.
\end{equation}
we find
\begin{eqnarray}\notag
&&(R')^2\left(\frac{f^2(u)v^2(r) r_i+2M-r_i}{r_i}  \right)-2R'Ef(u)v(r)\\
&-&\frac{\sin^2\theta (\partial_{\theta} R)^2-r^2(E^2-m^2)\sin^2\theta+j^2}{r^2\sin^2\theta}=0
\end{eqnarray}

Solving this equation we find 
\begin{equation}\label{20}
R_{\pm}(r)=\int\frac{E f(u)v(r) r_i \pm \sqrt{E^2r_i^2 \left(1-\frac{2M}{r_i}\right)-\Delta \,\mathcal{N} }}{r_i \Delta(r) } \mathrm{d}r,
\end{equation}
with $\Delta$ and $\mathcal{N}$ are given by  Eqs. (20) and (21).  Specializing the WH (BH) solution we need to set $f(u)=\pm 1$, which leads to the same conclusions as in the case of vector field.

\section{Tunneling with PW method
}
The basic idea behind this method is to apply the radial null geodesics which can be found by the metric \eqref{1}. In our case we are left with the following result
\begin{equation}
\dot{r}=f(u)v(r)\pm \sqrt{1-\frac{2M}{r_i}},
\end{equation}
in which the $+(-)$ gives the outgoing (ingoing) geodesics. The tunneling rate is related to the imaginary part of the action in the classically forbidden
region. In our paper the black hole mass is held fixed and the total ADM mass allowed to vary. When a shell of energy $\omega$ tunnels from the black hole, $M$ should be replaced by $M-\omega$. The imaginary part of the action is written as
\begin{equation}
\text{Im}S=\text{Im}\int_{r_{i}}^{r_{f}}p_r dr=\text{Im}\int_{r_{i}}^{r_{f}}\int_{0}^{p_r} p^{\prime}_r \mathrm{d}r
\end{equation}

In the black hole case [$f(u)=-1$] we find
\begin{eqnarray}
\text{Im}S&=&\text{Im}\int_{M}^{M-\omega}\int_{r_i}^{r_{f}}\frac{\mathrm{d}r}{\dot{r}}dH\\\notag
&=&\text{Im}\int_{0}^{\omega}\int_{r_i}^{r_{f}}\frac{\mathrm{d}r}{-v(r)+ \sqrt{1-\frac{2(M-\omega^{\prime})}{r_i}}}\mathrm{d}(-\omega^{\prime}).
\end{eqnarray}

Note that the integral becomes singular at the horizon. In particular we need to choose a positive sign which corresponds to the outgoing particles. Note that to ensure the positive result we deform the a semi-circle to give $-\pi i\, \text{Res}[f(x)]$, after solving this integral we end up with the following result
\begin{eqnarray}
\text{Im}S&=& \text{Im}\int_{0}^{\omega}4 \pi\,i (M-\omega^{\prime})\sqrt{1-\frac{2(M-\omega^{\prime})}{r_i}} \mathrm{d}\omega^{\prime}\\\notag
&\simeq & 4 \pi \omega \left(M-\frac{\omega}{2}\right)-\frac{4 \pi}{r_i}\left(M^2\omega-M \omega^2+\frac{\omega^3}{3}+...   \right) 
\end{eqnarray}

The above result can be approximated linear to $\omega$ as follows
\begin{eqnarray}
\text{Im}S&\simeq & 4 \pi \omega M \left(1-\frac{M}{r_i}\right),
\end{eqnarray}
with the tunneling rate
\begin{equation}
\Gamma_{BH}=\exp\left(-2\text{Im}S\right)=\exp\left[-8 \pi \omega M \left(1-\frac{M}{r_i}\right) \right].
\end{equation}

Making use of the Boltzmann equation $[\Gamma_{BH}=\exp(-\omega/T_{H}]$ we find
\begin{equation}
T_{H}=\frac{1}{8 \pi M} \left(1+\frac{M}{r_i}+...\right).
\end{equation}

Considering the fact that $r_i>> M $, we end up with the same equation 
\begin{equation}
T_{H}=\frac{\kappa_{BH}}{2\pi}=\frac{\mathcal{F}^{\prime}(r_h)}{4 \pi}.
\end{equation}

In the white hole case [$f(u)=1$], when we apply the same procedure we find 
\begin{eqnarray}
\text{Im}S&=&\text{Im}\int_{M}^{M-\omega}\int_{r_i}^{r_{f}}\frac{\mathrm{d}r}{\dot{r}}\mathrm{d}H\\\notag
&=&\text{Im}\int_{0}^{\omega}\int_{r_i}^{r_{f}}\frac{\mathrm{d}r}{v(r)- \sqrt{1-\frac{2(M-\omega^{\prime})}{r_i}}}\mathrm{d}(-\omega^{\prime}).
\end{eqnarray}

Note that a minus sign which corresponds to the ingoing particles has been chosen. To ensure the positive result we deform the a semi-circle to give $\pi i\, \text{Res}[f(x)]$ which gives 
\begin{eqnarray}
\text{Im}S&=& \text{Im}\int_{0}^{\omega}4 \pi\,i (M-\omega^{\prime})\sqrt{1-\frac{2(M-\omega^{\prime})}{r_i}} \mathrm{d}\omega^{\prime}\\\notag
&\simeq & +4 \pi \omega \left(M-\frac{\omega}{2}\right)-\frac{4 \pi}{r_i}\left(M^2\omega-M \omega^2+\frac{\omega^3}{3}+...   \right) 
\end{eqnarray}

Considering only the terms linear in $\omega$, this result can be approximated as 
\begin{eqnarray}
\text{Im}S&\simeq & 4 \pi \omega M \left(1-\frac{M}{r_i}\right),
\end{eqnarray}
with the tunneling rate
\begin{equation}
\Gamma_{WH}=\exp\left(-2\text{Im}S\right)=\exp\left[-8 \pi \omega M \left(1-\frac{M}{r_i}\right) \right].
\end{equation}

Finally approximating the solution when $r_i>>M$, yields
\begin{equation}
T_{H}=-\frac{\kappa_{WH}}{2\pi}=\frac{\mathcal{F}^{\prime}(r_h)}{4 \pi}.
\end{equation}

\section{Conclusion}

In this letter we have considered the Hawking radiation associated to the BH/WH geometry. We have used a generalized Painlev\'{e} coordinates together with the WKB and HJ methods. We have shown that a Hawking radiation is associated to the WH spacetime. Although particles can tunnel from the outside to the inside, based on the infinite valued flow analogue model one can argue that they will be spitted out again implying that only the outgoing mode is of primary interest.  Based on this, we have shown that the Hawking temperature outside the WH is the same as the BH temperature. Furthermore we have verified our result using the PW method. It is worth noting that, besides the mass, HT is affected by the initial radial distance $r_i$ at which the gravitational collapse starts. In general, HR can be considered as a negligible effect, but it remains to be seen if this effect could have any impact on the BHWH transition process. Finally we wish to point out that quantum gravity effects may effect this picture, in particular one may incorporate the GUP effects during the BH-to-WH transition, and see whether the infnite valued flow can really disappear. We plan in the near future to study the problem of Hawking radiation from acoustic WHs in the tunneling approach with GUP effects.

\section*{Acknowledgement}
The author would like to thank the editor and the referees for their valuable comments which helped to improve the manuscript.

\end{document}